\documentstyle[prd,aps]{revtex}

\begin{document}

\draft

\title{Large quantum gravity effects and nonlocal variables}
\author {Alfredo E. Dominguez\thanks{Electronic address: domingue@fis.uncor.edu} \and Manuel H. Tiglio\thanks{Electronic address: tiglio@fis.uncor.edu} }
\address{\it Facultad de Matem\'{a}tica, Astronom\'{\i}a y F\'{\i}sica,
Universidad Nacional de C\'{o}rdoba \\ Ciudad Universitaria 
5000 C\'{o}rdoba, Argentina. }
\maketitle

\begin{abstract}
We reconsider here the model where large quantum gravity effects were first found, but now in its Null Surface Formulation (NSF). We find that although the set of coherent states for $Z$, the basic variable of NSF, is as restricted as it is the one for the metric, while some type of small deviations from these states may cause huge fluctuations on the metric, the corresponding fluctuations on $Z$ remain small. 

\end{abstract}

 
\section*{}

An interesting effect found recently, that points out to potential
problems and/or surprises that could be found in quantum gravity, is the presence, in the quantization of certain midisuperspaces, of what has been called ``large quantum gravity effects" (LE). In the original analysis of Ashtekar \cite{ash1}, the model chosen is rotationally symmetric $2+1$ gravity coupled to a massless scalar field (a spacetime that may be obtained by symmetry reducing Einstein-Rosen waves). As shown in \cite{ash1}, in the quantized version of this model, coherent states for the scalar field emerge as natural candidates for a set of coherent states for the metric. It is then noted that only at low frequencies the mean value of the metric in these states is sharply peaked around its classical value. This implies that there are classical solutions that are spurious: they do not arise as classical limits of the quantum theory. And, curiously, even states very close to those that minimize the uncertainties on the metric cause huge fluctuations on the later. On the other hand, Gambini and Pullin
 have analyzed a different family of states, for which the uncertainties on the metric can be diminished at the expense of increased fluctuations on the matter fields \cite{gam}. Beetle \cite{bee} and Varadarajan \cite{var} have also analyzed other models, along lines similar to those of Ashtekar, and found results in agreement to those in \cite{ash1}.

Although the usual formulation of General Relativity (GR) is in terms of local variables, e.g., metric, connection, etc., there exists a reformulation, the Null Surface Formulation of GR (NSF), whose basic variable can be taken as a $``Z''$ function, that, {\em a posteriori} turns out to describe light cone cuts at future null infinity of a metric that satisfies Einstein equations. The purpose of this paper is to reconsider the model analyzed by Ashtekar, but now focusing our attention, not on the metric, but on $Z$. As we shall see, coherent states for the scalar field are also coherent states for $Z$, in exactly the same regime as that for the metric. But, while small deviations from this regime are amplified in the metric, they are damped in $Z$. A recent review of NSF, together with a complete list of references, can be found in \cite{koz}, and some results of classical NSF in $2+1$ in \cite{tan}.

Let us start by briefly discussing some classical aspects of the model. Coordinates (``Weyl coordinates'') can be chosen, on which the scalar field $\Psi$ decouples from the metric. The equations for $\Psi$ are those of a scalar field on a flat fiducial spacetime. Once $\Psi$ is known, the metric can be obtained by quadratures, and it takes the form
$$
ds^2 = e^{G\Gamma (r,t)} (-dt^2 +dr^2) +r^2 d\phi ^2 \;\;\;\;\; , \;\;\;\;\; \Gamma (r,t) \equiv \frac{1}{2} \int_0^r \left[ \left( \frac{\partial \Phi (\tilde{r},t) }{\partial \tilde{r} } \right) ^2 + \left( \frac{\partial \Phi (\tilde{r},t) }{\partial t} \right )^2 \right] \tilde{r}d\tilde{r} \; .
$$
Note that $\Gamma (r,t)$ is the energy of $\Psi$ enclosed on a disk of radius $r$, at time $t$, just as would be measured if the fiducial flat spacetime were physical. The range for the coordinates is the usual: $t \in {\cal R}, r \geq 0 , \phi \in [0, 2\pi ]$. The total mass is $H = \frac{1}{4G}(1 - e^{-4GH_0} ) $, and it coincides with the total energy of the scalar field, the ``C-energy'' $H_0 \equiv \Gamma (\infty ,t)$, only in the linearized approximation (in particular, $H$ is bounded from above, as opposed to $H_0$) \cite{ash2}. 

Suppose from now on that the initial data (at, say, $t=0$) for the scalar field has compact support, contained in $r\leq R_0$, and call ${\cal O}$ the region  outside the support of the field, i.e., the region given by $t\geq 0$ y $t-r\equiv u < -R_0$. In ${\cal O}$ the metric is
\begin{equation}
ds^2 = e^{GH_0} (-dt^2 +dr^2) +r^2 d\phi ^2  \; .  \label{met}
\end{equation}
It is in this zone and for the metric (\ref{met}) that the LE have been found.

Consider the light cone emanating from a point $x^{\mu} = (t,r,\phi)$, and its intersection with future null infinity ${\cal I}^+$ \cite{ash3}. This ``light cone cut''  \cite{cut} is a closed curve (with winding number $1$ and, in general, with selfintersections and cusps), that can be described by an equation of the type $u = Z(\xi, x^{\mu}, [\Psi])$, where $(u,\xi)$ are Bondi coordinates for ${\cal I ^+}$, and $[\Psi]$ denotes functional dependence with $\Psi$. Conversely, for each $(u,\xi) \in {\cal I}^+$, the set of points $x^{\mu}$ that satisfy $u = \mbox{constant} = Z(\xi, x^{\mu}, [\Psi]) $ comprise the past light cone of $(u,\xi)$. In this way, $Z$ admits two dual interpretations. 

If a $Z(\xi, x^{\mu} ,[\Psi])$ function satisfies the so called metricity conditions, then there exists a conformal metric such that  $Z$ has the interpretation above mentioned. This metric can be obtained as follows. For each $\xi$, there exist intrinsic coordinates $\theta ^{\nu} = (u , \omega, R)$, given by
$$
u = Z(\xi, x^{\mu} ,[\Psi]) \;\;\;\;\; , \;\;\;\;\; \omega = \partial _{\xi} Z(\xi, x^{\mu} ,[\Psi]) \;\;\;\;\; , \;\;\;\;\; R = \partial ^2 _{\xi} Z(\xi, x^{\mu} ,[\Psi]) \; .
$$
The $\Lambda (\xi , \theta ^{\nu},[\Psi])$ function is defined by expressing the coordinates $\theta ^{\nu}$ in term of $x^{\mu}$, inserting them in $R$, and differentiating with respect to $\xi$:
$$
\Lambda (\xi, \theta ^{\nu}, [\Psi]) \equiv \partial  _{\xi} R(\xi, x^{\mu} (\theta ^{\nu}) ,[\Psi]) \; .
$$
The cobasis $\theta^{\nu} _ a \equiv (d\theta ^{\nu})_a$, and its dual $\theta_{\mu}^a \theta ^{\nu}_a  = \delta ^{\;\; \nu } _ {\mu}$, are introduced. Then the components of the conformal metric in the basis $\{ \theta_{\mu}^a \}$ are
\begin{equation}
g^{ij} = \tilde{ \Omega } 
\left(
\begin{array}{ccc}
0 & 0 & -1 \\
0 & -1 & -(1/3) \partial_R \Lambda \\
-1 \;\;\; &  -(1/3)\partial_R \Lambda  \;\;\; & [-(1/3) \partial _{\xi} ( \partial_R  \Lambda ) + (1/9) ( \partial_R \Lambda )^2 + \partial_{\omega} \Lambda  ] 
\end{array}
\right) \; . \label{conf}
\end{equation}
One must also solve certain equations that ensure that $\tilde{\Omega }$ is such that (\ref{conf}) satisfies Einstein's equations.

In order to make contact with \cite{ash1}, we also look at the region ${\cal O}$. We then restrict ourselves to the portion of ${\cal I}^+$ given by $u \leq -R_0$ and $\xi <\phi  + arcos\left( t/r+R_0/r \right)$, where $Z$ is given by
\begin{equation}
Z(\xi, x^{\mu}, [\Psi]) = t - r \cos{[e^{-GH_0/2}(\xi - \phi)]} \; .  \label{cut}
\end{equation}
The main reason for such restriction is that it suffices to obtain the metric at $(t,r,\phi ) \in {\cal O}$. That is, following the procedure above sketched one finds that 
\begin{equation}
ds^2 =  \tilde{\Omega }^{-1} e^{-2GH_0}  \left[ e^{GH_0} (-dt^2 +dr^2) +r^2 d\phi ^2 \right] \; .
\label{metric}
\end{equation}
Instead of solving the corresponding equations for $\tilde{\Omega }$, we note that if we choose it as an arbitrary constant, then (\ref{metric}) satisfies the vacuum equations. We can rescale it as $\tilde{\Omega }= e^{-2GH_0}$, to reobtain in this way the metric in Weyl coordinates. Once this is done, it can be checked that, fixing a point $(t,r,\phi ) \in {\cal O}$, (\ref{cut}) gives the portion of the light cone cut that, on its way to ${\cal I}^+$, did not cross matter.

Summarizing, at this point we have two equivalent descriptions of the geometry in the region ${\cal O}$. One of them is given by the local variable (\ref{met}), and the other by the nonlocal one (\ref{cut}).

In the usual approach one encodes the degrees of freedom in $\Psi$, promotes it to an operator acting on the Hilbert space suggested by the fiducial flat background and obtains the covariant metric operator \cite{ash4} 
$$
\widehat{ds^2} = e^{G\widehat{H_0}} (-dt^2 + dr^2) + r^2 \widehat{I} d\phi ^2 \; ,
$$
whose unique non trivial component is $g \equiv e^{GH_0}$. Within NSF, we can quantize the light cone cuts, by promoting $Z$ to an operator ($\partial _{\xi} Z$ and $\partial ^2_{\xi} Z$ also have to be considered, but the results are similar)  \cite{fuz}:
$$
\widehat{Z} = Z (\xi, x^{\mu}, [ \hat{\Psi } ]) = t \widehat{I} - r \cos{[ e^{-G\widehat{H_0}/2}(\xi - \phi)]} \; .
$$

If the coherent states for the scalar field (for simplicity we consider monochromatic states, of frequency $\omega _0$):
\begin{equation}
|\Psi > = e^{-|c_0|^2/2\hbar }\sum_{n\geq 0} \left( \frac{c_0^2}{\hbar} \right )^{n/2} \frac{1}{\sqrt{n! }} |n_{\omega _0}>   \label{state} \;.
\end{equation}
are taken as candidates for coherent states for the metric, closed expressions for its classical value, $g \equiv e^{G<H_0>}$, as well as for its mean value and fluctuations, can be given. They are:
\begin{equation}
g  =  e^{a_0}  \;\;\;\;\; , \;\;\;\;\; 
<\widehat{g}>  =  e^{a_0 \left( e^{\Omega _0} -1\right) / \Omega _0} \;\;\;\;\; , \;\;\;\;
\left( \frac{\Delta \widehat{g}}{<\widehat{g}>} \right)^2  =  e^{a_0 \left(e^{\Omega _0} -1 \right) ^2} - 1 \; . \label{mop}
\end{equation}
with $N_0 \equiv |c_0|^2 / \hbar \gg1$ the number of particles, $\Omega _0 \equiv G\hbar \omega _0$, and $a_0 \equiv N_0 \Omega _0$. The condition $N_0 \gg 1$ must then be satisfied, in order for the uncertainties on the scalar field to be small. But this does not suffice for the mean value of $\widehat{g}$ to be sharply peaked around $g$; from eqs. (\ref{mop}) it can be seen that, for fixed but arbitrary $a_0$, this happens only at low frequencies, $\Omega _0 \ll 1$. In such a case, one can obtain the quantum corrections in a perturbative expansion around $\Omega _0 = 0$. Although we do not have closed expressions for the mean value of the cut and its fluctuations (the last are straightforward to obtain but rather lengthy, so we do not write them down hereafter), exactly the same above mentioned properties are valid for the cut: its mean value,
$$
<\widehat{Z}>  =  t - r \sum_{n\geq 0} \frac{(\xi - \phi)^{2n}}{(2n)!}(-1)^n e^{a_0(e^{-\Omega _0n}-1)/\Omega _0} \; ,
$$
is sharply peaked around the classical one, 
$$
Z = t  - r \cos{[ e^{-a_0/2}(\xi - \phi)]} \label{clas}  \; ,
$$
only at low frequencies. In that case, one can obtain the quantum corrections in a perturbative expansion:
\begin{eqnarray*}
<\widehat{Z}>& = & Z +  \frac{1}{2} a_0 \Omega _0\partial _{a_0^2} Z  + O\left( \Omega _0 ^2 \right)  \\
\left( \frac{\Delta \widehat{Z}}{<\widehat{Z}>} \right)^2 & = & \frac{1}{4Z^2} r^2 (\xi - \phi)^2 \sin ^2{[ e^{-a_0/2}(\xi - \phi)]} a_0e^{-a_0} \Omega _0+ O\left( \Omega _0^2 \right)   \; .
\end{eqnarray*}
Therefore, up to this point we do not find differences between metric and cut. But now suppose that certain number $N_1$ of particles of high frequencies is added to the state given by  (\ref{state}), with $N_0 \ll 1$, $\Omega _0 \ll 1$ and fixed but arbitrary $a_0$ (i.e., a coherent state for the scalar field that also minimizes the uncertainties for the metric, or, equivalently, for the cut) , i.e.,
\begin{equation}
|\Psi > = e^{- \left( |c_0|^2 + |c_1|^2 \right) /2\hbar }\sum_{n\geq 0} \sum_{m\geq 0} \left( \frac{c_0^2}{\hbar} \right )^{n/2} \left( \frac{c_1^2}{\hbar} \right )^{m/2} \frac{1}{\sqrt{m!n! }} |n_{\omega _0} m_{\omega _1}> \label{state2} \;.
\end{equation}
with $\Omega _1 \equiv G\hbar \omega _1 \gg1 $, $N_1 \equiv |c_1|^2 / \hbar $ (the number of particles added) and fixed but arbitrary $a_1\equiv N_1 \Omega _1$. Closed expressions for the metric quantities can also be given:
\begin{equation}
g  =  e^{a_0+a_1}  \; , \; 
<\widehat{g}>  =  e^{a_0 \left( e^{\Omega _0} -1\right) / \Omega _0 + a_1 \left( e^{\Omega _1} -1\right) / \Omega _1} \; , \;
\left( \frac{\Delta \widehat{g}}{<\widehat{g}>} \right)^2  =  e^{a_0 \left(e^{\Omega _0} -1 \right) ^2/\Omega _0 + a_1 \left(e^{\Omega _1} -1 \right) ^2/\Omega _1} - 1 \; .
\end{equation}
Recalling that $\Omega _0 \ll 1 $ and $\Omega _1 \gg 1$, we have that
\begin{equation}
<\widehat{g}>  \approx  e^{a_0 + a_1 e^{\Omega _1} / \Omega _1} \;\;\;\;\; , \;\;\;\;
\left( \frac{\Delta \widehat{g}}{<\widehat{g}>} \right)^2  \approx  e^{a_1 e^{2\Omega _1} /\Omega _1} - 1 \; .
\end{equation}
Now suppose that the number of particles with high frequency is very small, $N_1 \ll 1$, such that $a_1 \ll1$. This corresponds to a small deviation from the regime in which the pair $(\widehat{\Phi}, \widehat{g})$ is sharply peaked in their corresponding classical values (a ``little blip''). Even if $a_1$ is very small, for $\Omega _1$ large enough, the fluctuations on  $\widehat{g}$ are huge, its mean value differs in the same way from the classical one. The terms of the form $e^{a_1 e^{\Omega _1} /\Omega _1}$ cause this behavior. At this point is where the cut behaves in a completely different way. It can be seen that its fluctuations are very small, in fact there are terms of the form $\left( e^{a_1 e^{-\Omega _1} /\Omega _1} -1\right)$, and they quickly to zero when $\Omega _1$ increases. In other words, while small fluctuations are ``amplified'' in the metric, they are ``damped'' on the cut. The mean value of the cut has similar behavior, it remains very close to the classical value. The later is 
$$
Z = t - r \cos{\left[ e^{-(a_0 +a_1)/2}(\xi - \phi) \right] } \mbox{ , and therefore } Z \approx t - r \cos{\left[ e^{-a_0 /2}(\xi - \phi) \right] }
\mbox{ for } a_1 \ll 1 \; ;
$$
while the mean value is given by
$$
<\widehat{Z}> =  t - r \sum_{n\geq 0} \frac{(\xi - \phi)^{2n}}{(2n)!}(-1)^n e^{a_0(e^{-\Omega _0n}-1)/\Omega _0 + a_1e^{-\Omega _1n}-1)/\Omega _1}  \; , 
$$
and therefore, for $ \Omega _0 \ll1$, $a_1 \ll 1$, and  $\Omega _1 \gg1$, we have
$$
<\widehat{Z}> \approx  t - r \sum_{n\geq 0} \frac{(\xi - \phi)^{2n}}{(2n)!}(-1)^n e^{a_0 n}  = t - r \cos{\left[ e^{-a_0 /2}(\xi - \phi) \right] } \; .
$$

In conclusion, if the coherent states for the whole theory are taken as the ones that minimize the uncertainties on the scalar field, then they have to satisfy $N_0 \gg1$ and $\Omega _0 \ll 1$, whether the geometry is characterized by the metric or the cuts. If these conditions are not satisfied, the classical values of the cut and metric are spurious, in the sense given above. Nevertheless, there is a significative difference between metric and cut: a ``little blip'' of high frequency causes the metric to have wild fluctuations and hugely deviates its mean value from the classical one; as opposed to this, the mean value of the cut remains very close to the classical one and fluctuations on it are damped. Finally, we mention that within NSF one can also quantize local variables corresponding to the spacetime points \cite{fuz} and it can be seen that they behave in a manner very similar to that of the metric (i.e., they have huge fluctuations, etc., details will be considered elsewhere). 

The fact that fluctuations in $Z$ are damped while the ones for the covariant metric are amplified can be understood noticing from (\ref{conf}) that the components of the contravariant metric typically go as (derivatives of) $Z$. Now, if fluctuations for the covariant metric are large, it is quite natural that the ones for its inverse are small. Thus, one expects that similar results may hold in $3+1$, where one has an expression similar to (\ref{met}); in the sense that one could try to see if large quantum gravity effects are present in that case by analyzing the fluctuations in $Z$ \footnote{We thank J. Pullin for pointing out all this to us.}. But whether or not the later are damped will depend on the relation between the free Bondi data at ${\cal I}^+$ and $Z$. The problem is that at present such relation is only known in perturbative schemes. Since LE are nonlinear effects, it does not seem possible to obtain information from a perturbative approach.

The authors thank A. Ashtekar, R. J. Gleiser, M. Iriondo, C. N. Kozameh, and J. Pullin for valuable discussions, specially AA and CNK for suggesting this work. MHT is particularly indebted with R. Omi. AED acknowledges financial support from CONICOR. MHT also acknowledges financial support from CONICOR and CONICET. This work was supported in part by funds of the University of C\'ordoba, and grants from CONICET and CONICOR (Argentina). 

\end{document}